# Insights into the high-pressure behaviour of solid bromine from hybrid DFT calculations

*Madhavi H. Dalsaniya,[1,2] Krzysztof Jan Kurzydłowski,[1] Dominik Kurzydłowski[2]\**

[1] *Faculty of Materials Science and Engineering, Warsaw University of Technology, Wołoska 141, 02-507, Warsaw, Poland*

[2] *Faculty of Mathematics and Natural Sciences, Cardinal Stefan Wyszyński University in Warsaw, 01-038 Warsaw, Poland*

*d.kurzydlowski@uksw.edu.pl*

Understanding the properties of molecular solids at high pressure is a key element in the development of new solid-state theories. However, the most commonly used generalized-gradient approximation (GGA) of the density functional theory (DFT) often fails to correctly describe the behavior of these systems at high pressure. Here we utilize the hybrid DFT approach to model the properties of elemental bromine at high pressure. The calculations reproduce in very good agreement with experiment the properties of the molecular phase I (*Cmca* symmetry) and its pressure-induced transition into the non-molecular phase II (*Immm*). The experimentally yet unobserved transition into phase III (*I4/mmm*) is predicted to occur at 128 GPa, followed by subsequent formation of an *fcc* lattice at 188 GPa. Analysis of the structure and electronic properties of the modelled phases indicates that the molecular *Cmca* phase becomes metallic just at the borderline of its stability, and that both *Immm* and *I4/mmm* phases are metallic and quasi-2D. Finally, we show that the incommensurate phases of bromine postulated from experiment are transient species which can be viewed as intermediates in the dissociation process occurring at the boundary of the transition from phase I to phase II.

## I. INTRODUCTION

High-pressure research provides fundamental insight into material properties, which is essential for the development and validation of new condensed matter and planetary models. [1,2] One of the ongoing efforts in this field is the study of the process of molecular dissociation under pressures in the range of hundreds of gigapascal. [3,4] In particular, solids consisting of diatomic molecules, such as $H_2$, $N_2$, $O_2$ and the halogens ($F_2$, $Cl_2$, $Br_2$, $I_2$), have been studied at high pressure. Hydrogen, the first element in the periodic table, hosts within its molecule the simplest covalent bond. Theoretical considerations indicated that pressure-induced dissociation of this bond and formation of an alkali-metal type structure would lead to room-temperature superconductivity. [5,6] Despite the many theoretical and experimental studies devoted to this system the pressure of hydrogen dissociation and metallization is still not established. [7–13] Nitrogen is abundant in nature forming an exceptionally strongly bonded diatomic molecule. However, at high pressure and temperature (>110 GPa and >2,000 K), the triple N≡N intramolecular bond transforms into three N-N single bonds leading to the formation of a polymeric structure, which has been proposed as a high-energy-density material. [14–16] In contrast to both $H_2$ and $N_2$, the molecular crystal of oxygen is predicted to transform into a polymeric spiral chain structure only at an extremely high pressure of 1920 GPa. [17–19]

In general, high pressure leads to a reduction in volume, and a decrease in mean interatomic distances. This leads to the modification of material properties which as shown by the abovementioned examples can be surprisingly different for seemingly similar molecular crystals. Halogens ($F_2$, $Cl_2$, $Br_2$, $I_2$) are another representative of diatomic molecules whose high-pressure behaviour serves as an important analogue for both hydrogens, as they undergo pressure-induced metallization and dissociation, and oxygen, as their molecules are electron-rich species. Importantly, the respective phase transitions are known to take place in halogens at much lower pressures than for both hydrogen and oxygen. This is particularly true for the heavier halogens, $Br_2$ and $I_2$, which have been the most extensively studied, [20–25] while solids of $F_2$ and $Cl_2$ have gained relatively less attention. [26–29]

Iodine is solid at ambient conditions and forms a molecular crystal of *Cmca* symmetry and 8 atoms (four molecules) in the unit cell. Previous research showed that there is no structural phase transition in this element up to 20.6 GPa when metallization occurs. [25] A later study by Takemura et al. [21] reported molecular dissociation in iodine at 21 GPa when it transforms to a non-molecular orthorhombic structure (phase II) of *Immm* symmetry. Further experiments revealed that at 43 GPa the *Immm* structure transforms to a tetragonal phase III of *I4/mmm* symmetry ($Z = 2$). Above 55 GPa iodine enters phase IV with a closed-packed *fcc* structure (*Fm$\bar{3}$m*). [30] In addition to these phases two incommensurate structures, phase V and VI, have been found in the pressure range 22–26 GPa, that is in the vicinity of the *Cmca – Immm* phase boundary. [31,32] The emergence of these structures has been linked to the appearance of new Raman band in solid iodine at 23.5 GPa. [33,34]

The high-pressure phase transition sequence of bromine is similar to that of iodine, although the phase transitions are shifted to higher pressure. Bromine is liquid at room temperature and atmospheric pressure. It crystallizes at 266 K into a *Cmca* structure analogous to that of phase I of iodine. [35] Synchrotron X-ray diffraction studies revealed that solid bromine begins to undergo the *Cmca* to *Immm* phase at 80±5 GPa. [36,37] Raman scattering experiments were also performed on compressed solid bromine yielding frequency-pressure dependences for the intramolecular vibrational $A_g$ and $B_{3g}$ modes, as well as the intermolecular librational $A_g$ and $B_{3g}$ modes. [35,38] As in the case of iodine incommensurate structures of bromine were postulated to emerge at the *Cmca* to *Immm* transition. An incommensurate phase (V) was proposed by T. Kume at. al [33] based on the appearance at 84 GPa of two new Raman modes ($A_g^{(L)}$ and the X band) whose frequency decreases with increasing pressure. This softening was previously considered to be related to the onset of molecular dissociation as a displacive structural phase transition, and was also observed in the incommensurate phase for iodine. [16] Synchrotron X-ray data presented by Liu et al. seems to confirm the existence of phase V at 81 GPa, although the authors did not provide any structural details of this phase. [39] According to the previous electrical conductivity measurements, bromine metallization occurs at 60-70 GPa when it is still in the molecular *Cmca* phase. [40]

*Ab initio* modelling, mostly within the framework of Density Functional Theory (DFT), is an important tool for inspiring and guiding high pressure experiments, as well as aiding data interpretation. Several theoretical studies for iodine and bromine have been reported using the plane-wave pseudopotential approach with the generalized gradient approximation (GGA). [41–45] Based on calculations utilizing the PBE functional with the D2 van der Waals correction, Wu et al. [44] proposed that the splitting of the $A_g$ and $B_{3g}$ intramolecular Raman bands observed in compressed bromine in the 25 – 60 GPa pressure range [33] is due to symmetry lowering of the molecular phase from *Cmca* to *C2/m*. Their results indicated that bromine should metallize at 42.5 GPa while still in the molecular phase. Recently, four commensurate approximations for the incommensurate phase V of bromine were proposed based on DFT calculations. [45]

Although considerable research effort has been put into studying the high-pressure phase transitions of solid halogens, many questions still remain opened. As noted by Dalladay-Simpson et al. [28] large discrepancies exist in the experimental phase boundary pressures of halogens and those modelled with DFT. This might signal that the commonly used GGA approximation might be insufficient to correctly describe the properties of these systems. Indeed, a need for including more advanced functionals in modelling the high-pressure behaviour of hydrogen has been recently put forward. [12] Here we examined the phase transition sequence for solid elemental bromine from atmospheric pressure up to 200 GPa ($\approx 2 \cdot 10^6$ atm) by employing the hybrid HSE06 functional. We found that employed method correctly reproduces the phase stability of bromine both at ambient conditions and high pressure, in contrast to the GGA approximations used in previous theoretical studies. Apart from making reliable predictions on the yet unexplored phase transitions, our calculations offer insights into the evolution of the structural and electronic properties of compressed bromine, and the nature of the incommensurate phases of this solid.

## II. COMPUTATIONAL METHODS

Periodic-boundary conditions calculations utilizing the density functional theory (DFT) were used for investigating the geometry and enthalpy of solid bromine. The hybrid (HSE06), [46] meta-GGA (r$^2$SCAN), [47,48] and GGA (PBE) [49] functionals were utilized with appropriate van der Waals corrections (D3 method of Grimme), [50–52] as implemented the VASP 6.2 code [53,54]. The projector-augmented-wave method was adopted with $4s^2$ and $4p^5$ states treated as valence electrons.

All geometry parameters (lattice vectors and fractional coordinates) of the studied structures were optimized at selected pressures in the 0-200 GPa pressure range in 5 GPa intervals. The cut-off for plane wave energy was 800 eV. The convergence criterion for the electronic minimization was $10^{-7}$ eV. The Brillouin zone was probed through a Monkhorst-Pack mesh, with a $2\pi \times 0.033$ Å$^{-1}$ spacing of $k$- points. [55] All structures were optimized until the forces acting on the atoms were smaller than 1 meV/Å. Calculations of Γ-point vibration frequencies of the *Cmca* structure were conducted in VASP 6.2 utilizing the HSE06 functional. The finite-displacement method was used with a 0.025 Å displacement, and a tighter SCF convergence criterion ($10^{-8}$ eV). The HSE06 functional was also used to calculate the electronic band gap ($E_g$) of bromine with a $k$- points ($2\pi \times 0.025$ Å$^{-1}$). Visualization of all structures was performed with the VESTA. [56] For symmetry recognition we used the FINDSYM. [57] Group theory analysis of the vibrational modes was performed with the use of the Bilbao Crystallographic Server. [58] Bonding analysis was performed with the Lobster 4.1.0. [59–61]

## III. RESULTS AND DISCUSSION

### A. Modelling properties of solid bromine at ambient pressure

Although DFT methods are highly successfully at material modelling both at ambient and high pressure, in some cases the most commonly used generalized-gradient approximation (GGA) approximation does not correctly reproduce the actual properties. Hence, methods from higher rungs of the Jacob's ladder of DFT functionals, [62] such as meta-GGA functionals, are needed to correctly modelled such systems to avoid invoking empirical parameters such as the Hubbard U. [63,64] In the case of iodine and bromine it was found that both GGA and meta-GGA methods fail to describe correctly the ground state structure at 1 atm yielding a structure of *Cmcm* symmetry, composed of monoatomic chains, lower in energy than the experimentally observed molecular *Cmca* phase. [65] The correct energetic preference (*Cmca* lower than *Cmcm*) was obtained only with hybrid functionals.

This surprising result cast doubt on previous theoretical studies on the high-pressure behaviour of bromine, all of which used GGA methods.

In order to evaluate the influence of the functional on the resultant simulated properties of solid bromine we optimized the *Cmca* molecular structure at 1 atm using the three above-mentioned approximation: GGA (PBE functional), meta-GGA (r$^2$SCAN), and hybrid (HSE06) all with the D3 dispersion correction which is essential for the correct reproduction of the experimental geometry. The comparison of geometry parameters obtained with those functionals, given in Fig. 1, clearly shows an improved agreement between theory and experiment when moving from GGA through meta-GGA, to hybrid methods (see Table S1 in the Supplemental Material [66] for more details). The overestimation of the Br-Br bond length ($d_1$) drops from 4.2 % (PBE+D3) through 2.3 % (r$^2$SCAN+D3) to 0.3 % (HSE06+D3). The same trend is seen for intermolecular (van der Waals) contacts ($d_2$, $d_3$).

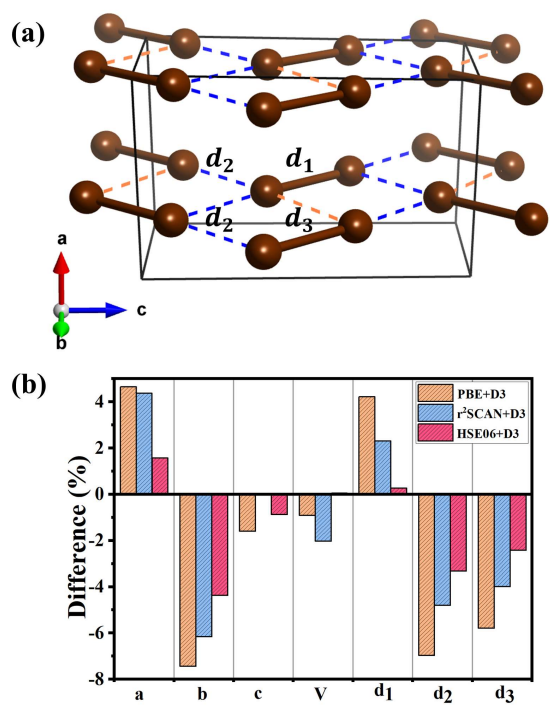

**FIG. 1** (a) Crystal structure of solid bromine in the *Cmca* molecular structure (phase I). (b) Difference between calculated and experimental geometry of the *Cmca* phase at 1 atm.

TABLE I Comparison between the calculated (this work) and experimental [67] properties of the bromine molecule: the dissociation energy (corrected for zero-point energy motion, $D_0$ in eV), the Br-Br bond distance ($r_1$ in Å), an its stretching frequency ($v$ in cm$^{-1}$). The energetic difference between the *Cmcm* and *Cmca* structures of solid bromine at 1 atm (in meV per atom) obtained with various DFT functionals in this work is also shown.

| Method | $D_0$ | $r_1$ | $v$ | $E_{Cmcm} - E_{Cmca}$ |
|---|---|---|---|---|
| PBE + D3 | 2.47 | 2.39 | 314 | –25 |
| r$^2$SCAN + D3 | 3.18 | 2.35 | 325 | –13 |
| HSE06 + D3 | 2.30 | 2.30 | 335 | +15 |
| Exp. | 1.97 | 2.28 | 325 | – |

The need for using the hybrid DFT method when modelling elemental bromine is corroborated by the comparison of the dissociation energy Br$_2$ molecule (Table 1). The GGA and meta-GGA functionals overestimate Br$_2$ stability, whereas the hybrid functional yields better agreement with experiment. In accordance with a previous study [65], we find that only the hybrid functional yields the experimental *Cmca* structure as lower in energy than the erroneous *Cmcm* structure which is composed of linear chains of bromine atoms (Table 1). Therefore, we choose this method for further calculations.

### B. Pressure stability of phases I-V

A variety of structures have been proposed for solid bromine under pressures. For the moment we will concentrate on the commensurate phases observed for this element (I – *Cmca* and II – *Immm*), as well as those found for iodine at higher pressure (III – *I4/mmm* and IV - *Fm$\bar{3}$m*). [43–45] The relative enthalpy of these phases (with respect to *I4/mmm*) calculated at the HSE06+D3 level of theory is shown in Fig.2. Our calculations indicate that the *Cmca* to *Immm* phase transition should take place at 90 GPa, in excellent agreement with the experimental value of 80±5 GPa. [36] Importantly, we reproduce the postulated first-order nature of the *Cmca* – *Immm* phase transition with a volume reduction of 2.4 %. This is in contrast with previous LDA and GGA modelling in which the *Cmca* structure transforms directly to the *Immm* structure upon geometry optimization at about

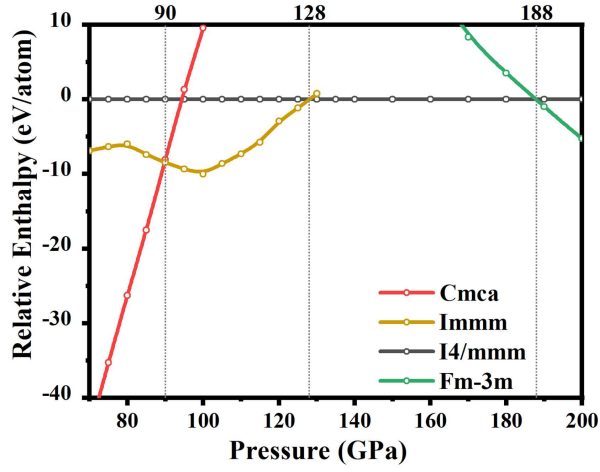

**FIG. 2** Pressure dependence of the enthalpies of high-pressure phases of bromine relative to the *I4/mmm* structure. Vertical lines denote the phase transition pressures (at 0 K).

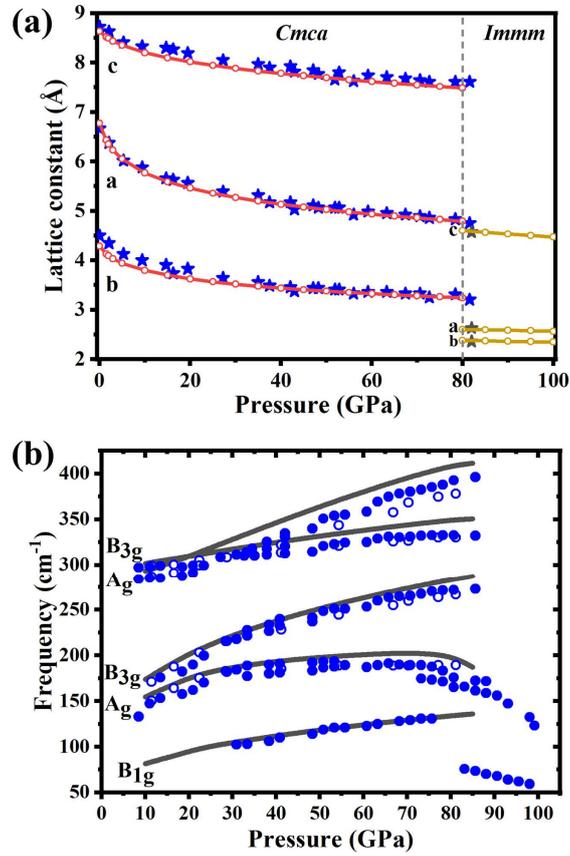

**FIG. 3** (a) Comparison of experimental lattice constants (blue/grey stars for *Cmca*/*Immm*, data taken from ref. [37]) with those calculated with the HSE06 functional (red/yellow lines for *Cmca*/*Immm*); (b) Comparison of experimental frequencies of Raman bands for the *Cmca* phase (blue close and open circles for data from ref. [33] and [38], respectively) and those calculated with HSE06 (grey line).

75 GPa. [42,43] We predict the yet unobserved transformations of *Immm* to the *I4/mmm* structure at 128 GPa, and the transition of this tetragonal structure to an *fcc* arrangement at 188 GPa. These transitions pressures are about 25 % higher than obtained from GGA modelling. [45]

Because our calculations correctly reproduce the high-pressure stability of the molecular *Cmca* phase, we are able to simulate the properties of this phase in the whole pressure range covered by experiment. As can be seen in Fig. 3a the theoretical lattice constants of the molecular *Cmca* phase are in excellent agreement with those obtained by x-ray diffraction, the same accordance is also found for the one pressure point for which experimental data was reported for the *Immm* structure. Moreover, the vibrational frequencies of the Raman-active bands of phase I obtained with the HSE06 functional agree very well with experimental data. Our results confirm the pressure-induced upshift of the frequencies of the intramolecular vibrations (high-frequency $A_g$ and $B_{3g}$ modes), as well as softening of the intermolecular $A_g$ mode above approximately 60 GPa. A previous study postulated that a splitting of the Raman bands of phase I in the 25 – 60 GPa pressure range [33] is related to a symmetry lowering in the molecular phase from *Cmca* to *C/2m*. [44] However we find that geometry optimization of the *C2/m* structure at various pressures always leads to the *Cmca* structure. The postulated symmetry lowering might be a result of nonhydrostatic conditions during the experimental measurements, or appearance of stacking faults in the *Cmca* phase.

### C. Chemical bonding and electronic properties of compressed bromine

The excellent agreement between our results and experiment lends credibility to further analysis of the properties of the high-pressure phases of bromine, in particular the chemical bonding picture which is not directly accessible through experiment. The *Cmca* phase, shown again in Fig. 4a, is predicted to retain its molecular structure up to the phase transition to *Immm*. Although considerable pressure-induced shortening of intermolecular contacts (marked $d_2$ and $d_3$, see Fig. 4a) is observed up to this pressure, they still remain more than 10 % longer than the intramolecular Br-Br bond (Fig. 4b). These contacts link the molecules into planes separated by an even larger distance ($d_{inter}$). We find good accordance of the calculated $d_1 - d_3$ distances with ambient pressure neutron diffraction data, [68] and high-pressure EXAFS experiments. [69] However we do not observe any considerable pressure-induced lengthening of the Br-Br bond ($d_1$ distance) as postulated from the latter. Comparison of EXAFS data with our calculations and the experimental ambient-pressure values (Fig. 4b) suggests that the lengthening observed in the EXAFS experiments might be an artefact resulting from the underestimation of the Br-Br bond length at pressure below 20 GPa.

The molecular character of solid bromine is lost upon the transition from *Cmca* to *Immm*, as evidenced by the near equalization of the intra- and intermolecular contacts (Fig. 4b). While in the former phase each Br atom forms one short bond ($d_1$ = 2.25 Å at 90 GPa) and three longer contacts (two at a distance

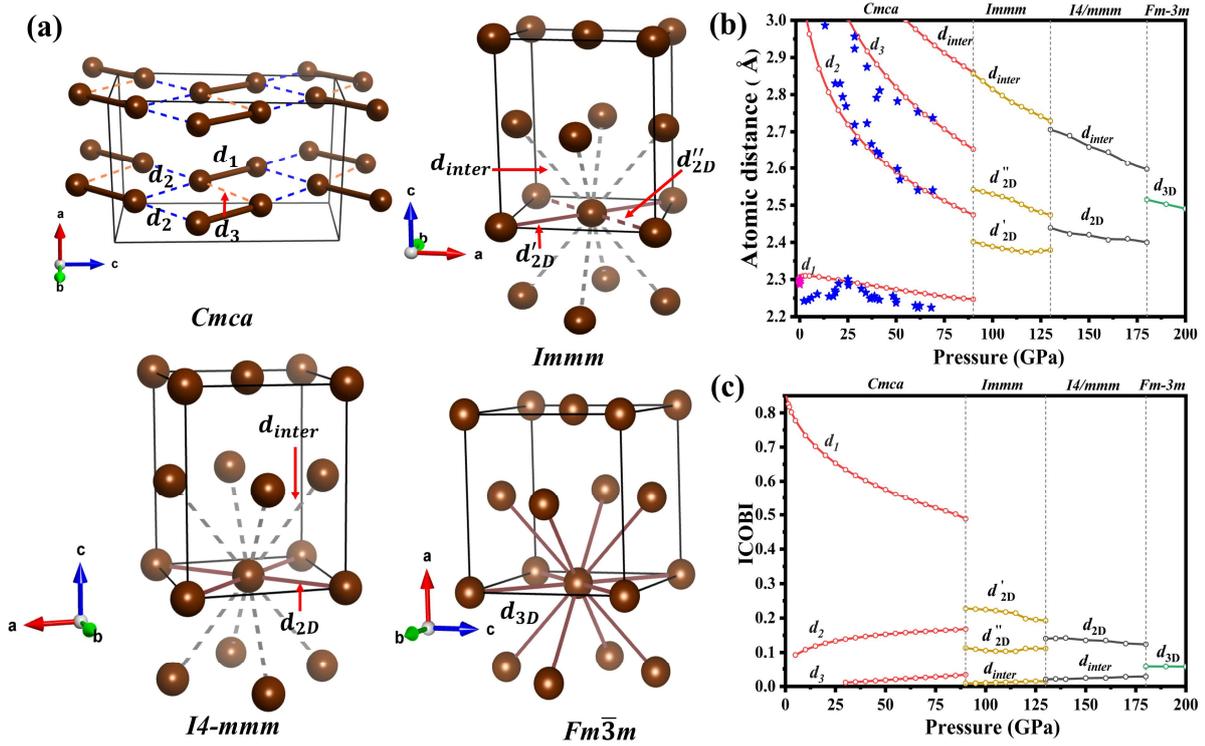

**FIG. 4** (a) The crystal structures of the high-pressure phases of solid bromine: phase I (*Cmca*), phase II (*Immm*), phase III (*I4/mmm*), and phase IV (*Fm$\bar{3}$m*); (b) Calculated pressure dependence of the Br-Br distances in these phases (red/yellow/grey/green lines for phases I/II/III/IV). Stars denote experimental values (blue – ref. [69] pink – ref. [68]) (c) Integrated crystal orbital bond index (ICOBI) for Br-Br distances. Dashed vertical lines in (b) and (c) indicate the predicted transition pressures between ground state structures of solid bromine.

$d_2$ = 2.47 Å and one at $d_3$ = 2.65 Å), in the latter two four short contacts ($d_{2D}'$ = 2.40 Å, $d_{2D}''$ = 2.54 Å) form a distorted square lattice (Fig. 4a). The *Immm* phase has a layered character, as the distance between planes formed by the $d_{2D}$ contacts ($d_{inter}$ = 2.86 Å at 0 GPa) is considerably larger than these contacts. Upon compression of *Immm* the difference between $d_{2D}'$ and $d_{2D}''$ decreases from 7 % at 90 GPa to 4 % at 128 GPa. At this pressure a phase transition to the *I4/mmm* structure is predicted, in which equalization of the two distances occurs, and hence a perfect square lattice (with $d_{2D}$ = 2.44 Å) is formed. The *I4/mmm* has still layered character as the difference between $d_{2D}$ and $d_{inter}$ is sizeable (11 % at 128 GPa). Finally, the transition to the *fcc* arrangement (the *Fm$\bar{3}$m* structure) marks equalization of $d_{2D}$ and $d_{inter}$ with the formation of a three-dimensional (3D) structure with each Br atom surrounded by 12 nearest-neighbours at $d_{3D}$ = 2.52 Å.

Further insights into the bonding in compressed bromine can be gained from the analysis of the recently introduced crystal orbital bond index (COBI), [60] which allows quantifying the strength of covalent interactions in solids, including those subject to high pressure. [70] The value of this index between a given pair of atoms integrated over occupied states (ICOBI) yields numbers corresponding to the bond order between those atoms. For the intramolecular Br-Br bond in *Cmca* at 1 atm the ICOBI value is 0.86, which corresponds well to the molecular picture of a molecule containing a single bond slightly destabilized by the antibonding effect of bromine lone pairs. The ICOBI values for $d_2$ and $d_3$ intermolecular interactions are close to zero, as expected for weak van der Waals interactions. However, as shown in Fig 4c, upon compression they increase, especially $d_2$, due to the decrease in the length of the interactions. At the same time the ICOBI value for the intramolecular bond decreases substantially (to 0.55 at 60 GPa), although the $d_1$ distance also becomes shorter. This trend might seem unexpected at first glance, but it can be explained by the fact that compression of the Br-Br bond induces an increase in the antibonding effect from the bromine lone pairs. This interpretation is corroborated by the analysis of ICOBI values decomposed into *s* and *p* orbital states (Fig. S2 in the Supplemental Material [66]), which indicates that the pressure-induced decrease of ICOBI for the Br-Br bond is associated mostly with the decreased bonding in the *p* manifold.

The ICOBI values for bromine atom pairs in the *Immm*, *I4/mmm*, and *Fm$\bar{3}$m* structures do not exceed 0.25 in the whole pressure range signalising lack of appreciable covalent bonding. Indeed, all of these structures are metallic in the pressure range at which they are thermodynamically stable. The electronic density of states at the Fermi level increases in the sequence from *Immm* (0.31 states/eV per atom at 120 GPa) through *I4/mmm* (0.32) to *Fm$\bar{3}$m* (0.44) indicating the bromine becomes a better metal upon increase of dimensionality.

An important question in the study of the high-pressure properties of halogens was whether the band gap closure occurs when these elements are still in the molecular *Cmca* phase. Although for iodine band gap closure was postulated to occur in this phase, [25,71,72] a recent re-analysis of the diffraction patterns suggested that metallization occurs after the molecular phase transform to an incommensurate structure. [32] In case of bromine, previous GGA calculations indicated band gap closure in the molecular *Cmca* phase at 42.5 GPa. [44] However, the GGA method is known to underestimate the band gap, while inclusion of the exact Hartree-Fock exchange in hybrid functionals yields larger band

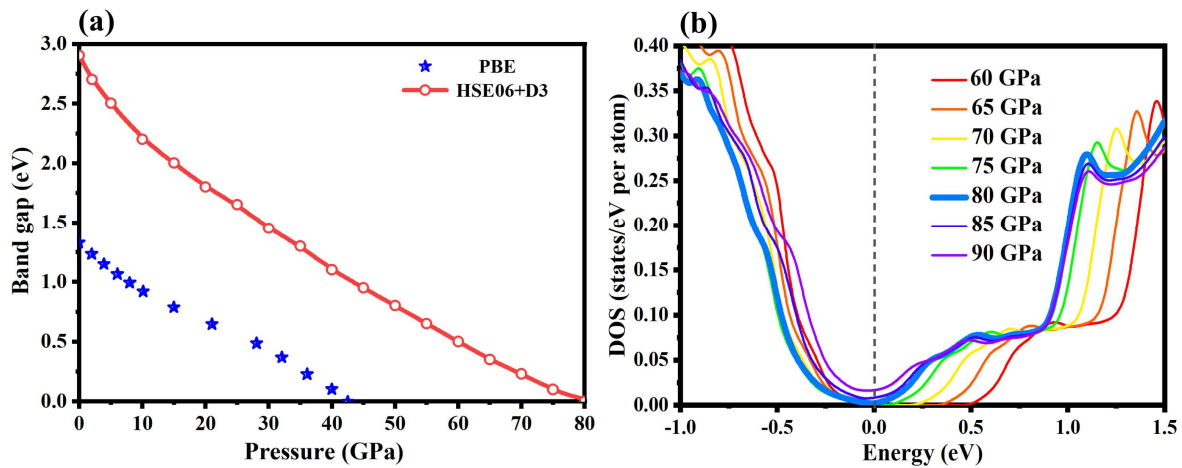

**FIG. 5** (a) Pressure dependence calculated for the *Cmca* phase of bromine with the HSE06+D3 method (red line) and previous results based on the PBE+D2 approach (blue stars, ref. [44]);(b) Electronic density of state near the Fermi level (indicated by a vertical line) for the *Cmca* phase of bromine between 60 and 90 GPa.

gaps which are in better agreement with experiment. [73] Indeed, our HSE06 calculation yield a much higher pressure of band gap closure of 80 GPa, as shown in Fig. 5. Therefore, our results show that metallization occurs while bromine is still in the molecular form. However, even after band gap closure the molecular phase remains a poor metal with the electronic density of states at the Fermi level below of 0.02 states/eV per atom).

### D. Incommensurate high-pressure phases of bromine

We now turn to the incommensurate structures which are observed on the verge of the *Cmca – Immm* phase transition. DFT modelling of these phases within periodic boundary conditions requires creating commensurate approximations, and a number of such proxies were proposed in previous studies. [27,43,45] These structures were described as derived from an *fco* arrangement of bromine atoms with their position modulated with a sine wave. We find two such modulated structures that are

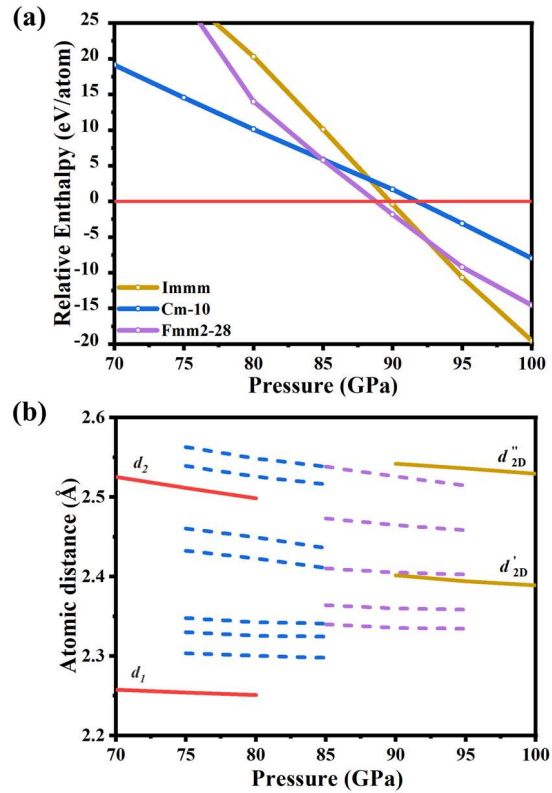

**FIG. 6** (a) Relative enthalpy of the *Immm, Cm-10, and Fmm2-28* structures with respect to *Cmca*. (b) Br-Br distances in the *Cmca/Immm* structures (red/yellow full lines) and *Cm-10/Fmm2-28* phases (blue/violet dashed lines).

competitive in terms of enthalpy with *Cmca* and *Immm*: the *Fmm2-28* phase (28 atoms in the unit cell) previously proposed by Li et al. [45], and a monoclinic *Cm-10* structure (10 atoms in the unit cell) which is a distorted variant of the *Fmm2-20* structure also described by these authors.

Previous GGA calculations indicated a large pressure range of stability (52 – 79 GPa) for the commensurate approximations of the incommensurate phases. [45] In contrast, our results, shown in Fig. 6 (a), give a very narrow stability range for *Fmm2-28* (89 – 92 GPa, just at the verge of the *Cmca – Immm* transition), while *Cm-10* is does not become the ground state structure at any pressure. However, the enthalpy differences between all four phases are less than 25 meV per atom (energy of room temperature thermal fluctuations) in the 70 – 100 GPa pressure range. Moreover, inclusion of corrections associated with the zero-point energy motion and vibrational entropy can further stabilize the modulated structures, as shown in the case compressed chlorine. [27] Evaluation of these corrections requires performing phonon calculations at various points of the Brillouin zone which was not possible given the large computational cost of the hybrid DFT approach. However, given the above one can conclude that the incommensurate phases can indeed be observed at the *Cmca – Immm* transition most probably due to stabilization by temperature effects. (it's noteworthy to point that all high-pressure experiments on bromine were performed at room temperature.) Furthermore, the analysis of Br-Br distances, shown in Fig. 6 (b), clearly shows that the subsequent transition from *Cmca* to *Cm-10* and then to *Fmm2-28* is connected with a stepwise elongation of the shortest Br-Br distance providing a path for the transformation from the molecular *Cmca* phase to the non-molecular and quasi-2D *Immm* structure. This progressive dissociation is connected with metallization of solid bromine in both modulated phases which are characterized by an electronic density of states at the Fermi level of 0.23/0.28 states/eV per atom for the *Cm-10*/*Fmm2-28* structures at 80 GPa, values substantially larger than found for the molecular *Cmca* structure.

## IV. CONCLUSIONS

The results obtained in this study indicate that in contrast to GGA the hybrid DFT method is capable of reproducing, with high accuracy, the behavior of compressed solid bromine. This enables the prediction of the yet unobserved transition from phase II (*Immm*) to III (*I4/mmm*) at 128 GPa, and subsequent formation of the *fcc* arrangement at 188 GPa, giving an incentive for future experiments. We find that the molecular *Cmca* phase becomes metallic at pressures higher than previously reported, [40] and that the incommensurate structures are indeed transient species which can be viewed intermediates in the dissociation process marked by the *Cmca* to *Immm* transformation. Partial dissociation of $Br_2$ molecules in those phases enhances the metallic properties. We show that both *Immm* and *I4/mmm* phases, while being metals, exhibit a layered structure. The formation of non-molecular phases can be attributed to the weakening of the Br-Br intramolecular bond through enhancement of the antibonding effect which stems from the pressure-induced shortening of this bond.

## ACKNOWLEDGMENTS

M. D. and D.K. acknowledges the support from the National Science Centre, Poland (NCN) within the SONATA BIS grant (no. UMO-2019/34/E/ST4/00445). This research was carried out with the support of the Interdisciplinary Centre for Mathematical and Computational Modelling at the University of Warsaw (ICM UW), under grant no. GA83-26. Part of the calculations were made at the Poznań Supercomputing and Networking Center (PSNC) within grant no. 562.

## REFERENCES

[1]   C. S. Yoo, *Chemistry under Extreme Conditions: Pressure Evolution of Chemical Bonding and Structure in Dense Solids*, Matter Radiat. Extremes **5**, 018202 (2020).

[2]   L. Zhang, Y. Wang, J. Lv, and Y. Ma, *Materials Discovery at High Pressures*, Nat. Rev. Mater.

**2**, 1 (2017).

[3]  W. Grochala, R. Hoffmann, J. Feng, and N. W. Ashcroft, *The Chemical Imagination at Work in Very Tight Places*, Angew. Chemie - Int. Ed. **46**, 3620 (2007).

[4]  R. J. Hemley, *Percy W. Bridgman's Second Century*, High Press. Res. **30**, 581 (2010).

[5]  E. Wigner and H. B. Huntington, *On the Possibility of a Metallic Modification of Hydrogen*, J. Chem. Phys. **3**, 764 (1935).

[6]  N. W. Ashcroft, *Metallic Hydrogen: A High-Temperature Superconductor?*, Phys. Rev. Lett. **21**, 1748 (1968).

[7]  J. M. McMahon and D. M. Ceperley, *Ground-State Structures of Atomic Metallic Hydrogen*, Phys. Rev. Lett. **106**, 165302 (2011).

[8]  R. T. Howie, C. L. Guillaume, T. Scheler, A. F. Goncharov, and E. Gregoryanz, *Mixed Molecular and Atomic Phase of Dense Hydrogen*, Phys. Rev. Lett. **108**, 125501 (2012).

[9]  C. Ji, B. Li, W. Liu, J. S. Smith, A. Majumdar, W. Luo, R. Ahuja, J. Shu, J. Wang, S. Sinogeikin, Y. Meng, V. B. Prakapenka, E. Greenberg, R. Xu, X. Huang, W. Yang, G. Shen, W. L. Mao, and H. K. Mao, *Ultrahigh-Pressure Isostructural Electronic Transitions in Hydrogen*, Nature **573**, 558 (2019).

[10] E. Gregoryanz, C. Ji, P. Dalladay-Simpson, B. Li, R. T. Howie, and H. K. Mao, *Everything You Always Wanted to Know about Metallic Hydrogen but Were Afraid to Ask*, Matter Radiat. Extremes **5**, 038101 (2020).

[11] M. I. Eremets, A. P. Drozdov, P. P. Kong, and H. Wang, *Semimetallic Molecular Hydrogen at Pressure above 350 GPa*, Nat. Phys. **15**, 1246 (2019).

[12] H. C. Yang, K. Liu, Z. Y. Lu, and H. Q. Lin, *First-Principles Study of Solid Hydrogen:*

*Comparison among Four Exchange-Correlation Functionals*, Phys. Rev. B **102**, 174109 (2020).

[13] P. Dalladay-Simpson, R. T. Howie, and E. Gregoryanz, *Evidence for a New Phase of Dense Hydrogen above 325 Gigapascals*, Nature **529**, 63 (2016).

[14] M. I. Eremets, R. J. Hemley, H. K. Mao, and E. Gregoryanz, *Semiconducting Non-Molecular Nitrogen up to 240 GPa and Its Low-Pressure Stability*, Nature **411**, 170 (2001).

[15] X. Wang, Y. Wang, M. Miao, X. Zhong, J. Lv, T. Cui, J. Li, L. Chen, C. J. Pickard, and Y. Ma, *Cagelike Diamondoid Nitrogen at High Pressures*, Phys. Rev. Lett. **109**, 175502 (2012).

[16] J. Sun, M. Martinez-Canales, D. D. Klug, C. J. Pickard, and R. J. Needs, *Stable All-Nitrogen Metallic Salt at Terapascal Pressures*, Phys. Rev. Lett. **111**, 175502 (2013).

[17] J. Sun, M. Martinez-Canales, D. D. Klug, C. J. Pickard, and R. J. Needs, *Persistence and Eventual Demise of Oxygen Molecules at Terapascal Pressures*, Phys. Rev. Lett. **108**, 045503 (2012).

[18] L. Zhu, Z. Wang, Y. Wang, G. Zou, H. K. Mao, and Y. Ma, *Spiral Chain $O_4$ Form of Dense Oxygen*, Proc. Natl. Acad. Sci. U. S. A. **109**, 751 (2012).

[19] B. H. Cogollo-Olivo, S. Biswas, S. Scandolo, and J. A. Montoya, *Phase Diagram of Oxygen at Extreme Pressure and Temperature Conditions: An Ab Initio Study*, Phys. Rev. B **98**, 094103 (2018).

[20] B. M. Riggleman and H. G. Drickamer, *Approach to the Metallic State as Obtained from Optical and Electrical Measurements*, J. Chem. Phys. **38**, 2721 (1963).

[21] K. Takemura, S. Minomura, O. Shimomura, and Y. Fujii, *Observation of Molecular Dissociation of Iodine at High Pressure by X-Ray Diffraction*, Phys. Rev. Lett. **45**, 1881 (1980).

[22] M. Pasternak, J. N. Farrell, and R. D. Taylor, *Metallization and Structural Transformation of Iodine under Pressure: A Microscopic View*, Phys. Rev. Lett. **58**, 575 (1987).

[23] H. Olijinyk, W. Li, and A. Wokaun, *High-Pressure Studies of solid Iodine by Raman Spectroscopy*, Phys. Rev. B **50**, 712 (1994).

[24] N. Sakai, K. Takemura, and K. Tsuji, *Electrical Properties of High-Pressure Metallic Modification of Iodine*, J. Phys. Soc. Japan **51**, 1811 (1982).

[25] C. Shimomura, K. Takemura, Y. Fujii, S. Minomura, M. Mori, Y. Noda, and Y. Yamada, *Structure Analysis of High-Pressure Metallic State of Iodine*, Phys. Rev. B **18**, 715 (1978).

[26] J. Shi, E. Fonda, S. Botti, M. A. L. Marques, T. Shinmei, T. Irifune, A.-M. Flank, P. Lagarde, A. Polian, J.-P. Itié, and A. San-Miguel, *Halogen Molecular Modifications at High Pressure: The Case of Iodine*, Phys. Chem. Chem. Phys. **23**, 3321 (2021).

[27] P. Li, G. Gao, and Y. Ma, *Modulated Structure and Molecular Dissociation of Solid Chlorine at High Pressures*, J. Chem. Phys. **137**, 064502 (2012).

[28] P. Dalladay-Simpson, J. Binns, M. Peña-Alvarez, M. E. Donnelly, E. Greenberg, V. Prakapenka, X. J. Chen, E. Gregoryanz, and R. T. Howie, *Band Gap Closure, Incommensurability and Molecular Dissociation of Dense Chlorine*, Nat. Commun. **10**, 1 (2019).

[29] D. Duan, Z. Liu, Z. Lin, H. Song, H. Xie, T. Cui, C. J. Pickard, and M. Miao, *Multistep Dissociation of Fluorine Molecules under Extreme Compression*, Phys. Rev. Lett. **126**, 225704 (2021).

[30] Y. Fujii, K. Hase, N. Hamaya, Y. Ohishi, A. Onodera, O. Shimomura, and K. Takemura, *Pressure-Induced Face-Centered-Cubic Phase of Monatomic Metallic Iodine*, Phys. Rev. Lett. **58**, 796 (1987).

[31] T. Kenichi, S. Kyoko, F. Hiroshi, and O. Mitsuko, *Modulated Structure of Solid Iodine during Its Molecular Dissociation under High Pressure*, Nature **423**, 971 (2003).

[32] H. Fujihisa, K. Takemura, M. Onoda, and Y. Gotoh, *Two Intermediate Incommensurate Phases in the Molecular Dissociation Process of Solid Iodine under High Pressure*, Phys. Rev. Res. **3**, 033174 (2021).

[33] T. Kume, T. Hiraoka, Y. Ohya, S. Sasaki, and H. Shimizu, *High Pressure Raman Study of Bromine and Iodine: Soft Phonon in the Incommensurate Phase*, Phys. Rev. Lett. **94**, 065506 (2005).

[34] Q. Zeng, Z. He, X. San, Y. Ma, F. Tian, T. Cui, B. Liu, G. Zou, and H. K. Mao, *A New Phase of Solid Iodine with Different Molecular Covalent Bonds*, Proc. Natl. Acad. Sci. U. S. A. **105**, 4999 (2008).

[35] P. G. Johannsen, and W. B. Holzapfel, *Effect of Pressure on Raman Spectra of Solid Bromine*, J. Phys. C: Solid State Phys. **16**, 1961 (1983).

[36] Y. Fujii, K. Hase, Y. Ohishi, H. Fujihisa, N. Hamaya, K. Takemura, O. Shimomura, T. Kikegawa, Y. Amemiya, and T. Matsushita, *Evidence for Molecular Dissociation in Bromine near 80 GPa*, Phys. Rev. Lett. **63**, 536 (1989).

[37] H. Fujihisa, Y. Fujii, K. Takemura, and O. Shimomura, *Structural Aspects of Dense Solid Halogens under High Pressure Studied by X-Ray Diffraction—Molecular Dissociation and Metallization*, J. Phys. Chem. Solids **56**, 1439 (1995).

[38] Y. Akahama, H. Kawamura, H. Fujihisa, K. Aoki, and Y. Fujii, *Raman Spectra of Solid Bromine under Pressure of up to 80 GPa*, Rev. High Press. Sci. Technol. **7**, 793 (1998).

[39] M. K. Liu, D. F. Duan, Y. P. Huang, Y. F. Liang, X. L. Huang, and T. Cui, *Reexploration of Structural Changes in Element Bromine through Pressure-Induced Decomposition of Solid*

*HBr∗*, Chinese Phys. Lett. **36**, (2019).

[40] K. Shimizu, K. Amaya, and S. Endo, *High Pressure Science and Technology: Proceedings of the Joint XV AIRAPT and XXXIII EHPRG International Conference, Warsaw, Poland, 1995 (World Scientific, Singapore, River Edge, NJ, 1996), p. 498.*, in (n.d.).

[41] H. Miyagi, K. Yamaguchi, H. Matsuo and K. Mukose, *First-Principles Study of Solid Iodine and Bromine under High Pressure*, J. Phys.: Condens. Matter **10**, 11203 (1998).

[42] M. S. Miao, V. E. Van Doren, and J. L. Martins, *Density-Functional Studies of High-Pressure Properties and Molecular Dissociations of Halogen Molecular Crystals*, Phys. Rev. B **68**, 094106 (2003).

[43] D. Duan, Y. Liu, Y. Ma, Z. Liu, T. Cui, B. Liu, and G. Zou, *Ab Initio Studies of Solid Bromine under High Pressure*, Phys. Rev. B **76**, 104113 (2007).

[44] M. Wu, J. S. Tse, and Y. Pan, *Anomalous Bond Length Behavior and a New Solid Phase of Bromine under Pressure*, Sci. Rep. **6**, 25649 (2016).

[45] P. Li, X. Du, G. Gao, R. Sun, L. Zhang, and W. Wang, *New Modulated Structures of Solid Bromine at High Pressure*, Comput. Mater. Sci. **171**, 109205 (2020).

[46] A. V. Krukau, O. A. Vydrov, A. F. Izmaylov, and G. E. Scuseria, *Influence of the Exchange Screening Parameter on the Performance of Screened Hybrid Functionals*, J. Chem. Phys. **125**, 224106 (2006).

[47] J. Sun, A. Ruzsinszky, and J. P. Perdew, *Strongly Constrained and Appropriately Normed Semilocal Density Functional*, Phys. Rev. Lett. **115**, 036402 (2015).

[48] J. W. Furness, A. D. Kaplan, J. Ning, J. P. Perdew, and J. Sun, *Accurate and Numerically Efficient $r^2$SCAN Meta-Generalized Gradient Approximation*, J. Phys. Chem. Lett. **11**, 8208

(2020).

[49] J. P. Perdew, K. Burke, and M. Ernzerhof, *Generalized Gradient Approximation Made Simple*, Phys. Rev. Lett. **77**, 3865 (1996).

[50] S. Grimme, J. Antony, S. Ehrlich, and H. Krieg, *A Consistent and Accurate Ab Initio Parametrization of Density Functional Dispersion Correction (DFT-D) for the 94 Elements H-Pu*, J. Chem. Phys. **132**, 154104 (2010).

[51] S. Ehlert, U. Huniar, J. Ning, J. W. Furness, J. Sun, A. D. Kaplan, J. P. Perdew, and J. G. Brandenburg, *r$^2$SCAN-D4: Dispersion Corrected Meta-Generalized Gradient Approximation for General Chemical Applications*, J. Chem. Phys. **154**, 061101 (2021).

[52] J. Moellmann and S. Grimme, *DFT-D3 Study of Some Molecular Crystals*, J. Phys. Chem. C **118**, 7615 (2014).

[53] G. Kresse and D. Joubert, *From Ultrasoft Pseudopotentials to the Projector Augmented-Wave Method*, Phys. Rev. B **59**, 1758 (1999).

[54] G. Kresse and J. Furthmüller, *Efficient Iterative Schemes for Ab Initio Total-Energy Calculations Using a Plane-Wave Basis Set*, Phys. Rev. B **54**, 11169 (1996).

[55] H. J. Monkhorst and J. D. Pack, *Special Points for Brillouin-Zone Integrations*, Phys. Rev. B **13**, 5188 (1976).

[56] K. Momma and F. Izumi, *VESTA 3 for Three-Dimensional Visualization of Crystal, Volumetric and Morphology Data*, J. Appl. Cryst. **44**, 1272 (2011).

[57] H. T. Stokes and D. M. Hatch, *FINDSYM : Program for Identifying the Space-Group Symmetry of a Crystal*, J. Appl. Cryst. **38**, 237 (2005).

[58] E. Kroumova, M. L. Aroyo, J. M. Perez-Mato, A. Kirov, C. Capillas, S. Ivantchev, and H.

Wondratschek, *Bilbao Crystallographic Server: Useful Databases and Tools for Phase-Transition Studies*, Phase Transitions **76**, 155 (2003).

[59] R. Nelson, C. Ertural, J. George, V. L. Deringer, G. Hautier, and R. Dronskowski, *LOBSTER: Local Orbital Projections, Atomic Charges, and Chemical-Bonding Analysis from Projector-Augmented-Wave-Based Density-Functional Theory*, J. Comput. Chem. **41**, 1931 (2020).

[60] P. C. Müller, C. Ertural, J. Hempelmann, and R. Dronskowski, *Crystal Orbital Bond Index: Covalent Bond Orders in Solids*, J. Phys. Chem. C **125**, 7959 (2021).

[61] R. Dronskowski and P. E. Blöchl, *Crystal Orbital Hamilton Populations (COHP). Energy-Resolved Visualization of Chemical Bonding in Solids Based on Density-Functional Calculations*, J. Phys. Chem. **97**, 8617 (1993).

[62] J. P. Perdew and K. Schmidt, *Jacob's Ladder of Density Functional Approximations for the Exchange-Correlation Energy*, AIP Conference Proceedings **577**, 1 (2001).

[63] C. Lane, J. W. Furness, I. G. Buda, Y. Zhang, R. S. Markiewicz, B. Barbiellini, J. Sun, and A. Bansil, *Antiferromagnetic Ground State of $La_2CuO_4$: A Parameter-Free Ab Initio Description Christopher*, Phys. Rev. B **98**, 125140 (2018).

[64] Y. Zhang, J. Furness, R. Zhang, Z. Wang, A. Zunger, and J. Sun, *Symmetry-Breaking Polymorphous Descriptions for Correlated Materials without Interelectronic U*, Phys. Rev. B **102**, 045112 (2020).

[65] J. George, C. Reimann, V. L. Deringer, T. Bredow, and R. Dronskowski, *On the DFT Ground State of Crystalline Bromine and Iodine*, ChemPhysChem **16**, 728 (2015).

[66] See Supplemental Material at [URL] for the calculated and experimental structural parameter of bromine at 1 atm; the electronic density of states, negative values of the COHP and COBI for the *Cmca* phase at 0 and 60 GPa; pressure dependence of the Boltzmann distribution between

*Cmca*, *Immm*, *Cm-10*, and *Fmm2-28* phases of bromine; lattice constants and fractional coordinates of the studied phases of bromine.

[67] K. P. Huber and G. Herzberg, *Constants of diatomic molecules. In: Molecular Spectra and Molecular Structure. Springer, Boston, MA.*, **1999** , *8-689 (1979)*.

[68] B. M. Powell, K. M. Heal, and B. H. Torrie, *The Temperature Dependence of the Crystal Structures of the Solid Halogens, Bromine and Chlorine*, Molecular Physics **53**, 929 (1984).

[69] A. San-Miguel, H. Libotte, M. Gauthier, G. Aquilanti, S. Pascarelli, and J. P. Gaspard, *New Phase Transition of Solid Bromine under High Pressure*, Phys. Rev. Lett. **99**, 015501 (2007).

[70] X. Wang, T. Bi, K. P. Hilleke, A. Lamichhane, R. J. Hemley, and E. Zurek, *Dilute Carbon in $H_3S$ under Pressure*, npj Comput Mater **8**, 87 (2022).

[71] T. Ishikawa, K. Mukai, Y. Tanaka, M. Sakata, Y. Nakamoto, T. Matsuoka, K. Shimizu, and Y. Ohishi, *Metallization of Solid Iodine in Phase I: X-Ray Diffraction Measurements, Electrical Resistance Measurements, and Ab Initio Calculations*, High Press. Res. **33**, 186 (2013).

[72] H. Nagara, T. Ishikawa, and T. Kotani, *Band Structure and Pressure-Induced Metallic Transition in Iodine - GW Calculation*, High Press. Res. **34**, 215 (2014).

[73] H. Xiao, J. Tahir-Kheli, and W. A. Goddard III, *Accurate Band Gaps for Semiconductors from Density Functional Theory*, J. Phys. Chem. Lett. **2**, 212 (2011).